\documentclass[aps,prb,superscriptaddress,twocolumn]{revtex4-1}
\usepackage{bm}
\usepackage[fleqn]{amsmath}
\usepackage{hyperref}
\usepackage{graphicx}

\begin{document}

\title{Resonant Enhancement of Second Harmonic Generation by Edge States in Transition Metal Dichalcogenide Monolayers}

\author{V.V. Enaldiev}
\email{vova.enaldiev@gmail.com}
\affiliation{Kotelnikov Institute of Radio-engineering and Electronics of the Russian Academy of Sciences, 11-7 Mokhovaya St, Moscow, 125009 Russia}
\affiliation{Laboratory of 2D Materials' Optoelectronics,  Moscow Institute of Physics and Technology, Dolgoprudny 141700, Russia}

\date{\today}

\begin{abstract}

We derive a low-energy theory for edge states in transition metal dichalcogenide  monolayers for a two-band $\bm{kp}$-Hamiltonian in case of uncoupled valleys. In the absence of spin-orbit interaction at the edge, these states possess a linear dispersion described by a single phenomenological parameter characterizing the edge structure. Depending on the sign of the parameter, the edge state spectrum can either cross the band gap or lie outside of it.  In the first case, the presence of edge states leads to resonant enhancement of the second harmonic generation at frequencies about the half of the band gap, in agreement with recent experiments. The value of the phenomenological boundary parameter is extracted from the resonance frequency position.

\end{abstract}

\maketitle


The optical properties of monolayer crystals of transition metal dichalcogenides (TMDs) (like MoS$_2$, MoSe$_2$, MoTe$_2$, WS$_2$, and WSe$_2$) have recently attracted a considerable interest due to the possible optoelectronic applications\cite{bib:Sun,bib:Posp}. This is due to the direct band gap of the monolayer TMDs whose value corresponds to the visible and infrared light frequencies\cite{bib:Falko}. The optical response of the bulk materials at the absorption edge is dominated by excitons\cite{bib:Moody}. However, recent experiment\cite{bib:Yin} have also demonstrated resonant enhancement of non-linear response near the edge of MoS$_2$ monolayer membranes at frequencies about the half of the band gap. This resonance was attributed with the existence of the edge states (ESs). 

The metallic ESs in the band gap have been lately observed in a MoS$_2$ monolayer on graphite using scanning tunnelling microscopy and spectroscopy\cite{bib:Zhang}. The dependence of ES properties in MoS$_2$ monolayer nanoribbons on the type of edge termination, passivation, and reconstruction were studied using density functional theory\cite{bib:Boll,bib:Ataca,bib:Erdogan,bib:Vojv,bib:Li} (DFT) as well as tight-binding approximation\cite{bib:Khoeini,bib:Liu} (TBA). However, description of the ESs in the TMD monolayers within the $\bm{kp}$-approach allow one to describe the ESs without going into details of microscopic structure of the edge and to take into account effects of external fields. This enables to construct an analytic theory for the ESs in the whole class of materials in a unified way. Such a general theory relies on a boundary condition (BC) that describes the edge structure by means of several phenomenological parameters\cite{bib:Korman}. The values of these parameters can be obtained by fitting with the experimental data or other calculations based on DFT or TBA. Recently, the ES spectra in the TMD monolayer nanoribbon\cite{bib:Segarra} and optical absorption in TMD nanoflakes involving transitions between the bulk and edge states\cite{bib:Trushin} have been studied in the $\bm{kp}$-approximation. However, these studies were restricted by some certain values of the phenomenological parameters.
 
The aim of this communication is to construct an analytical theory for the ESs in TMD monolayers in the $\bm{kp}$-approach and to reveal the effects of ESs on the non-linear optical response. We demonstrate that the ESs possess linear spectra which are described by a single real phenomenological parameter (for each spin value) in the absence of valley coupling and spin-orbit interaction at the edge. Sign of the parameter determines whether the ES spectra intersect the band gap or not. We show that the second harmonic generation is resonantly enhanced by the ESs crossing the band gap, and extract the value of the parameter from the experimentally observed\cite{bib:Yin} resonance frequency.

In the TMD monolayers the conductance and valence band edges are located in the $K$ and $K'$ valleys of the honeycomb lattice. Within a two-band $\bm{kp}$-approach, dynamics of electrons with spin $s/2=\pm 1/2$ in the $K(K')$ valley is described by the Hamiltonian\cite{bib:Falko,bib:Xiao}
\begin{equation}\label{TMD_Hamilonian}
H_{\tau, s} = \left( 
\begin{array}{cc}
m + s\tau\Delta_c & v\left(\tau p_x - ip_y\right) \\
v\left(\tau p_x + ip_y\right) & -m + s\tau\Delta_v
\end{array}
\right)
\end{equation}
where $2m$ is the band gap without spin splitting, $2\Delta_{c,v}$ is the value of spin splitting in the conduction and valence band correspondingly, the index $\tau=+1(-1)$ denotes the $K(K')$ valley,  $\bm{p}=\left( p_x, p_y \right)$ is the in-plane 2D momentum, $v$ is the velocity matrix element between the band extrema. The Hamiltonian $H_{\tau, s}$ (\ref{TMD_Hamilonian}) acts on the two-component wave function $\psi_{s,\tau} = (\psi_{c,s,\tau}, \psi_{v,s,\tau})^{T}$. As it was mentioned above, to describe the edge of the TMD monolayer one should supplement the Hamiltonian (\ref{TMD_Hamilonian}) with a BC for $\psi_{s,\tau}$. Here we consider only the zigzag or reconstructed zigzag types of edges for which projections of the valley centers onto the edge direction are well distant from each other. In this case we can neglect by the valley coupling at the edge. We also suppose that spin-orbit interaction is absent at the edge, so we can independently treat the two spin components in the Hamiltonian (\ref{TMD_Hamilonian}). Under this conditions the most general BC that entangles the components of the function $\psi_{s,\tau}$ is the same as in a single valley approximation of graphene \cite{bib:VVolkov,bib:Ostaay,bib:Zag}:  
\begin{equation}\label{BC}
\left[\psi_{c,s,\tau} + ia_{s,\tau}e^{-i\tau\phi}\psi_{v,s,\tau}\right ]_{\mbox{{\it at\, edge}}} = 0,
\end{equation}
where $a_{s,\tau}$ is the real phenomenological parameter for the valley $\tau$ and the spin $s/2$, characterizing the edge structure (including passivation, relaxation or reconstraction of the zigzag edge), $\phi$ is an angle between the vector of unit normal to the edge and the $x$-axis. The BC (\ref{BC}) is derived from vanishing of the probability current normal to the edge. The identity between the intravalley BC (\ref{BC}) for TMD monolayers and that for graphene stems from the equality of the current operator for the Hamiltonian (\ref{TMD_Hamilonian}) and that for graphene. Below we will consider the edges which preserve the time reversal symmetry. This symmetry imposes the following relation between the phenomenological parameters in the two valleys for the opposite spin values: $a_{s,\tau} =-a_{-s,-\tau}$. Here we note that study of Ref.[\onlinecite{bib:Trushin}] was limited by the BC (\ref{BC}) with $a_{1,1}=a_{-1,1}=-a_{1,-1}=-a_{-1,-1} = 1$. As the valleys and spins do not entangle in our consideration, the indexes $\tau, s$ will be hereafter suppressed  everywhere except where they are needed. 

\begin{figure}
\includegraphics[width=8cm,height=8cm]{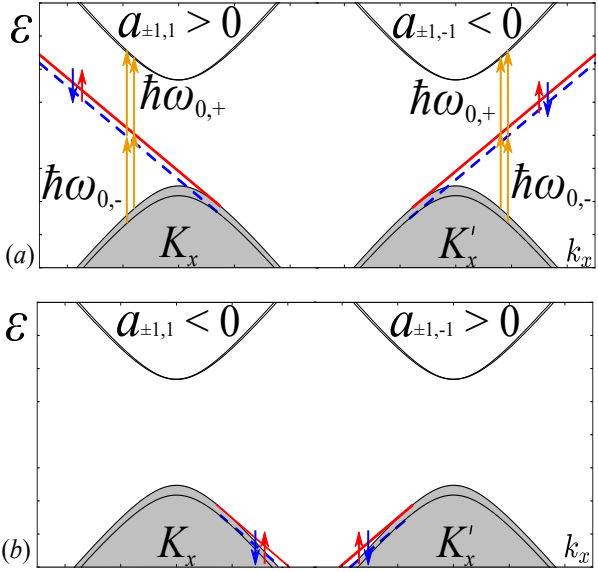}
\caption{\label{Fig:spectra} Energy spectra $\varepsilon(k_x)$ of the TMD monolayer half-plane in absence of valley coupling and spin-orbit interaction at the edge. Here, $k_x$ is the wave vector along the edge measured from the center of the edge Brillouin zone. Shaded regions correspond to the filled bulk states. Red solid ($s=1$) and blue dashed ($s=-1$) rays represent spectra of the edge states for the boundary parameters (a) $a_{1,1} = a_{-1,1}>0$, $a_{1,-1} = a_{-1,-1}<0$; (b) $a_{1,1} = a_{-1,1}<0$, $a_{1,-1} = a_{-1,-1}>0$. In case (a) orange arrows indicate transitions via the edge states that lead to the resonance in second harmonic generation at the frequencies $\omega_{0,\pm}$ (\ref{resonance_frequency}). }
\end{figure}

Now we consider the monolayer occupying the half-plane \text{$y>0$} with translation invariant edge characterized by constant $a_{s,\tau}$ along the $x$-axis. In this case, the quasimomentum component $p_x$ measured from the projections of valley centers on the edge is a good quantum number. The ES wave function that obeys the equation $H_{\tau,s}\psi_e = \varepsilon_e\psi_e$ and the BC (\ref{BC}) is
\begin{equation}\label{ES_function}
\psi_e = C_e
\left (
\begin{array}{c}
1 \\
-\frac{\tau}{a_{s,\tau}} 
\end{array}
\right )e^{-\kappa y + ik_xx}
\end{equation}
here we have introduced the 2D wave vector $\hbar\bm{k}=\bm{p}$, 
\begin{equation}
C_e = \left(\frac{2a_{s,\tau}^2\kappa}{L_x(1+a_{s,\tau}^2)(1-e^{-2\kappa L_y})}\right)^{1/2}
\end{equation} 
is the normalization factor, 
\begin{equation}\label{kappa}
\kappa = -\tau k_x\frac{1-a_{s,\tau}^2}{1+a_{s,\tau}^2} + \frac{m}{\hbar v}\frac{2\tau a_{s,\tau}}{1+a_{s,\tau}^2} - \frac{sa_{s,\tau}\left(\Delta_v - \Delta_c\right)}{\hbar v\left ( 1 + a_{s,\tau}^2\right )}
\end{equation}
is an inverse decay length of the edge states. The dispersion law of the ESs is expressed as follows (see Fig.\ref{Fig:spectra}):
\begin{equation}\label{ES_spectrum}
\varepsilon_e = -\frac{2a_{s,\tau}}{1+a_{s,\tau}^2} vp_x - \frac{1-a_{s,\tau}^2}{1+a_{s,\tau}^2}m + \frac{s\tau\left ( \Delta_v + a_{s,\tau}^2\Delta_c\right)}{1+a_{s,\tau}^2}, \kappa >0.
\end{equation}
The condition $\kappa > 0$ determines the energy range for the existence of the ESs. In case $a_{\pm 1,1}>0$ ($a_{\pm 1,-1}<0$) the spectrum of ESs (\ref{ES_spectrum}) intersects the band gap and has the end point in the valence band for \text{$0<a_{\pm 1,1}<1$} (\text{$-1<a_{\pm 1,-1}<0$}) or in the conduction band for \text{$a_{\pm 1,1}>1$} ($a_{\pm 1,-1}<-1$) in the valley $K$($K'$). In opposite case $a_{\pm 1,1}<0$ ($a_{\pm 1,-1}>0$) energies of the ESs lies outside the band gap and overlap with the valence band for \text{$-1<a_{\pm 1,1}<0$} (\text{$0<a_{\pm 1,-1}<1$}) or conduction band for $a_{\pm 1,1}<-1$ ($a_{\pm 1,-1}>1$) in the valley $K$($K'$).  

Now we consider the bulk states. In the infinite monolayer the Hamiltonian (\ref{TMD_Hamilonian}) is diagonalized in the plane wave basis:
\begin{equation}\label{plane_wave_function}
\psi_{l,\bm{k}}=C_{b,l}
\left(
\begin{array}{c}
1 \\
\frac{\varepsilon_{l}-m-s\tau\Delta_c}{\hbar vk}e^{i\theta_{k}}
\end{array}
\right)e^{ik_xx+ik_yy}\equiv \phi_{k_x,k_y}e^{ik_xx+ik_yy},
\end{equation}
here $k^2=k_x^2 + k_y^2$, $e^{i\theta_k}=\left(\tau k_x+ik_y\right)/k$, $l = 1 (2)$ is the conduction (valence) band index, the normalization factor is expressed as follows:
\begin{equation}\label{bulk_constant}
C_{b,l} = \frac{1}{\sqrt{L_xL_y}}\frac{\hbar v k}{\sqrt{(\hbar v k)^2 + (\varepsilon_{l} - m -s\tau\Delta_c)^2}}.
\end{equation}
The last equality in (\ref{plane_wave_function}) is the definition of the two-component function $\phi_{k_x,k_y}$. Spectra of bulk states is expressed by the formula:
\begin{equation}\label{bulk_spectra}
\varepsilon_{1,2} =  \varepsilon_{0,s,\tau} \pm \sqrt{\widetilde{m}_{s,\tau}^2 + (\hbar vk)^2},
\end{equation}
where \text{$\varepsilon_{0,s,\tau}=s\tau\left(\Delta_v+\Delta_c\right)/2$}, \text{$\widetilde{m}_{s,\tau} = m - s\tau\left(\Delta_v - \Delta_c\right)/2$}.

In the half-plane, the wave functions (\ref{plane_wave_function}) do not satisfy the BC (\ref{BC}). Therefore, we will seek the bulk wave function as a sum of incident and reflected plane waves with identical wave vector component $k_x$:
\begin{equation}\label{bulk_function}
\psi_b = \frac{1}{\sqrt{2}}\left[\phi_{k_x,-k_y}e^{ik_xx -ik_yy} + R\phi_{k_x,k_y}e^{ik_xx +ik_yy}\right],
\end{equation}  
here $k_y=\left\{\left[\left(\varepsilon_l-\varepsilon_{0,s,\tau}\right)^2 - \widetilde{m}_{s,\tau}^2\right]/(\hbar v)^2 - k_x^2\right\}^{1/2}>0$. The reflection coefficient $R$ is determined by the BC (\ref{BC}) and reads as follows:
\begin{equation}
R = -\frac{\hbar vk + a\tau\left(\varepsilon_l - m - s\tau\Delta_c\right)e^{-i\theta_k}}{\hbar vk + a\tau\left(\varepsilon_l - m - s\tau\Delta_c\right)e^{i\theta_k}}.
\end{equation}
As a consequence of the probability current conservation we have $|R| = 1$. The wave functions of the edge (\ref{ES_function}) and bulk (\ref{bulk_function}) states form the complete set in the system under consideration. Here, we note that for the half-plane problem the mirror symmetry plane $x\to-x$ may also present. However, existence of the mirror plane does not impose any additional constriction for the boundary parameter $a_{s,\tau}$. This means that arbitrary observable quantity (for example, the second order conductivity) calculated in the basis (\ref{ES_function}), (\ref{bulk_function}) will keep the mirror symmetry plane restrictions even if the latter is really not present at the edge on the atomic scale.  


Now we turn to the calculation of the second order conductivity which is responsible for the second harmonic generation and show that the ESs lying in the gap (see Fig.{\ref{Fig:spectra}(a)) resonantly enhance the latter. 
For this aim we solve the quantum kinetic equation 
\begin{equation}\label{kinetic_equation}
i\hbar \frac{\partial\rho}{\partial t} = \left[H_{s\tau}-e\left(\varphi_{\bm{q}\omega}e^{i\bm{qr}-i(\omega +i0)t}+c.c.\right), \rho\right]
\end{equation} 
for the density matrix $\rho$. The electric field is determined through the spatial derivative of the potential $E(\bm{r},t)= -i\bm{q}\varphi_{\bm{q}\omega}e^{i\bm{qr}-i(\omega +i0)t} + c.c.$ Below, we are interested in the limit $\bm{q}\to 0$ that should be taken with a care as linear terms in $\bm{q}$ to be preserved. The solution of Eq.(\ref{kinetic_equation}) can be expanded in the series of the electric field amplitude:
\begin{equation}\label{density_matrix_series}
\rho = \rho^{(0)} + \left[\rho^{(1)}e^{-i\omega t} + c.c.\right] + \left[\rho^{(2)}e^{-i2\omega t} + c.c.\right],
\end{equation} 
where $\rho^{(0)}$ is the equilibrium density matrix, $\rho^{(1)}\propto E$, $\rho^{(2)}\propto E^2$. The second harmonic current is determined by the $\rho^{(2)}$. In the basis (\ref{ES_function}), (\ref{bulk_function}) matrix elements of $\rho^{(2)}$ are expressed as follows:
\begin{eqnarray}\label{second_density_matrix}
\rho^{(2)}_{\lambda'\lambda} = e^2\varphi_{\bm{q}\omega}^2\sum_{\lambda''}\frac{\langle\lambda'|e^{i\bm{qr}}|\lambda''\rangle\langle\lambda''|e^{i\bm{qr}}|\lambda\rangle}{\varepsilon_{\lambda} - \varepsilon_{\lambda'} + 2\hbar\omega + i0}\times\nonumber\\ 
\times\left[\frac{f_{\lambda} - f_{\lambda''}}{\varepsilon_{\lambda} - \varepsilon_{\lambda''} + \hbar\omega + i0} - \frac{f_{\lambda''} - f_{\lambda'}}{\varepsilon_{\lambda''} - \varepsilon_{\lambda'} + \hbar\omega + i0}\right],
\end{eqnarray}
here $\lambda$ is a composite index running over the bulk state quantum numbers $\left\{\epsilon_l, k_x\right\}$ and the edge state quantum number ${k_{x}}$, $\rho^{(0)}_{\lambda\lambda'}=f_{\lambda}\delta_{\lambda\lambda'}$ is the equilibrium density matrix whose diagonal elements are the Fermi-Dirac distribution function $f_{\lambda}$. 

Using Eq.(\ref{second_density_matrix}) one readily obtains the current density at the double frequency as follows:
\begin{equation}\label{current}
j^{(2)\alpha}_{2\omega} = \frac{-e}{L_xL_y}\sum_{s\tau\lambda\lambda'}v^{\alpha}_{\lambda\lambda'}\rho^{(2)}_{\lambda'\lambda},
\end{equation} 
where $v^{\alpha} = \partial H_{s\tau}/\hbar\partial k_{\alpha}$ is the velocity operator ($\alpha = (x,y)$). Equation (\ref{current}) allows us to express the non-linear conductivity as $\sigma_{\alpha\beta\gamma}=\sum_{s\tau}\sigma_{\alpha\beta\gamma}^{(s,\tau)}$, here $\sigma_{\alpha\beta\gamma}^{(s,\tau)}$ is a contribution to the conductivity from the electrons with the spin $s/2$ in the valley $\tau$. The resonant term in the current (\ref{current}) emerges when the intermediate (generally virtual) states $\lambda''$ in Eq.(\ref{second_density_matrix}) coincide with some real states. For the frequencies less than the bulk band gap, this can only occur when the state $\lambda'$ belongs to the conduction band bulk, the state $\lambda$ belongs to the valence band bulk, and $\lambda''$ corresponds to the ESs intersecting the band gap (see Fig.\ref{Fig:spectra}a). Below, we analyse only the resonance terms $\delta\sigma_{\alpha\beta\gamma}$ in the conductivity $\sigma_{\alpha\beta\gamma}$. The calculations presenting in detail in Supplemental Material\cite{bib:Support} show that the resonance arises only in ${\rm Re}\left(\delta\sigma_{xxx}^{(s,\tau)}\right), {\rm Im}\left(\delta\sigma_{yxx}^{(s,\tau)}\right)$. After some algebra one finds
\begin{equation}\label{conductivity_resonance_1}
\begin{aligned}
\displaystyle
{\rm Re}\left(\delta\sigma^{(s,\tau)}_{xxx}\right) = \frac{\tau e^3v^2}{4L_y(\hbar\omega)^2}\Theta(\omega-\omega_{0,s\tau})\times\qquad\qquad\qquad\qquad\\
\times\frac{\omega_{0,s\tau}\left[\omega - \scalebox{1.2}{$\frac{2|a_{s\tau}|}{1+a_{s\tau}^2}$}\omega_{0,s\tau}+\scalebox{1.2}{$\frac{\left(1-a_{s\tau}^2\right)\omega_{0,s\tau}^2}{2\left(1+a_{s\tau}^2\right)\omega}$}\right]^2}{\omega\left[\omega^2 - \omega_{0,s\tau}^2\right]^{3/2}}\times\\
\times\left[f\left(\varepsilon_{0,s,\tau}-\hbar\omega\right)-2f\left(\varepsilon_{0,s,\tau}\right)+f\left(\varepsilon_{0,s,\tau}+\hbar\omega\right)\right]\qquad
\end{aligned}
\end{equation}
\begin{equation}\label{conductivity_resonance_2}
\begin{aligned}
{\rm Im}\left(\delta\sigma^{(s,\tau)}_{yxx}\right) = \frac{e^3v^2\left(1+a_{s\tau}^2\right)}{4L_y\left|a_{s\tau}\right|(\hbar\omega)^2}\Theta(\omega-\omega_{0,s\tau})\times\qquad\qquad\qquad\qquad\\
\times\frac{\omega_{0,s\tau}^2\left[\omega - \scalebox{1.2}{$\frac{2|a_{s\tau}|}{1+a_{s\tau}^2}$}\omega_{0,s\tau}+\scalebox{1.2}{$\frac{\left(1-a_{s\tau}^2\right)\omega_{0,s\tau}^2}{2\left(1+a_{s\tau}^2\right)\omega}$}\right]^2}{\omega^2\left[\omega^2 - \omega_{0,s\tau}^2\right]^{3/2}}\times\qquad\\
\times\left[f\left(\varepsilon_{0,s,\tau}-\hbar\omega\right)-2f\left(\varepsilon_{0,s,\tau}\right)+f\left(\varepsilon_{0,s,\tau}+\hbar\omega\right)\right]\qquad\qquad
\end{aligned}
\end{equation}
where $\Theta\left(...\right)$ is the Heaviside step function. The resonance frequency in the above equations is determined by the formula:
\begin{equation}\label{resonance_frequency}
\qquad\qquad\hbar\omega_{0,s\tau} = \frac{1+a_{s\tau}^2}{2\left|a_{s\tau}\right|}\widetilde{m}_{s\tau}.
\end{equation}
From the Eq.(\ref{resonance_frequency}) it follows that there are two resonance frequencies characterized by the sign of the product $s\tau$. The time reversal symmetry relates the latter from the different valleys with opposite spins $\omega_{0,s\tau}=\omega_{0,-s-\tau}$. The resonance contributions (\ref{conductivity_resonance_1}),(\ref{conductivity_resonance_2}) as a function of frequency are plotted in Fig.(\ref{Fig:Re_sigma_xxx}). The power-law singularity $\left(\omega-\omega_0\right)^{-3/2}$ in Eqs.(\ref{conductivity_resonance_1}),(\ref{conductivity_resonance_2}) is provided by the density of bulk states which is proportional to $k_y^{-1}$ and the product of matrix elements $\langle\varepsilon_{1}'k_x'|e^{i\bm{qr}}|k_x''\rangle\langle k_x''|e^{i\bm{qr}}|\varepsilon_{2}k_x\rangle$ that is proportional\cite{bib:Support} to $k_y^{-2}$. Therefore, the main contribution to the resonance stems from the bulk electrons propagating along the edge with $k_y\approx 0$. The resonance strength is lessened as $a_{s,\tau}$ tends to unity and it vanishes at $a_{s,\tau}=1$ (see Fig.(\ref{Fig:Re_sigma_xxx})). This happens because in numerators of the Eqs.(\ref{conductivity_resonance_1}),(\ref{conductivity_resonance_2}) arises $\left(\omega - \omega_{0,s\tau}\right)^2$ at $a_{s,\tau}=1$. The system size $L_y$ enters explicitly in the denominators of Eqs.(\ref{conductivity_resonance_1}),(\ref{conductivity_resonance_2}) due to matrix elements between the bulk and edge states. This is not unusual as the contribution of transitions between the bulk band states via the intermediate ESs to the non-linear two-dimensional conductivity should tend to zero as the system size tends to infinity. Nevertheless, this contribution leads to the non-vanishing total 2D current (even if the system size tends to infinity) and is readily observable in spatially-resolved experiments.

We also note that the conductivity ${\rm Re}\left(\delta\sigma_{xxx}^{(s,\tau)}\right)$ is proportional to the valley index $\tau$ and, therefore, has opposite sign in the two valleys. Consequently, the total resonance term in the non-linear conductivity ${\rm Re}(\delta\sigma_{xxx})=\sum_{s\tau}{\rm Re}(\delta\sigma_{xxx}^{(s,\tau)})$ is nonzero if only electrons have different energy distributions in the two valleys. It is manifestation of the time reversal symmetry imposing the invariance under the mirror symmetry $x\to -x$, which leads to the relation $\sigma_{xxx}^{s\tau}=-\sigma_{xxx}^{-s-\tau}$ (if there is no imbalance in the valley populations). The previous relation ensures the total conductivity $\sigma_{xxx}$ to be zero.  

Now we extract the value of the boundary parameter $a_{s,\tau}$ from the recent experiment\citep{bib:Yin} reporting the resonant second harmonic generation near the edge of a MoS$_2$ single-layer membrane. The resonance wavelength $1300$nm corresponds to the energy $\hbar\omega_{res}=0.954$eV. The bulk energy gap of atomically thin MoS$_2$ layer derived from the absorption spectroscopy\cite{bib:Mak,bib:Zande} is $1.8$eV. The DFT calculations\cite{bib:Falko} provide us with the following spin splitting in the valence band $2\Delta_v=0.148$eV and in the conduction band $2\Delta_c=0.003$eV. Therefore, we have $\widetilde{m}_{1,1}=\widetilde{m}_{-1,-1}=0.902$eV, $\widetilde{m}_{-1,1}=\widetilde{m}_{1,-1}=0.974$eV. The solutions of Eq.(\ref{resonance_frequency}) with respect to unknown $a_{s,\tau}$ exist only if $\hbar\omega_{res}/\widetilde{m}_{s,\tau}>1$. For the bulk parameters given above it is true only for $\widetilde{m}_{1,1}$ ($\widetilde{m}_{-1,-1}$) in the valley $K(K')$. There are two non-equivalent solutions for the phenomenological parameter $a^{(1)}_{1,1}=0.71$ and $a^{(2)}_{1,1}=1.4$ (remind that $a_{-1,-1}=-a_{1,1}$). Although we do not have additional information to choose the definite one, we can observe from Fig.(\ref{Fig:Re_sigma_xxx}) that the resonance for  $a^{(1)}_{1,1}=0.71$ is more pronounced than that for  $a^{(2)}_{1,1}=1.4$. Therefore, it is more likely that the boundary parameter is $a^{(1)}_{1,1}=0.71$.

\begin{figure}
\includegraphics[width=8cm, height=12cm]{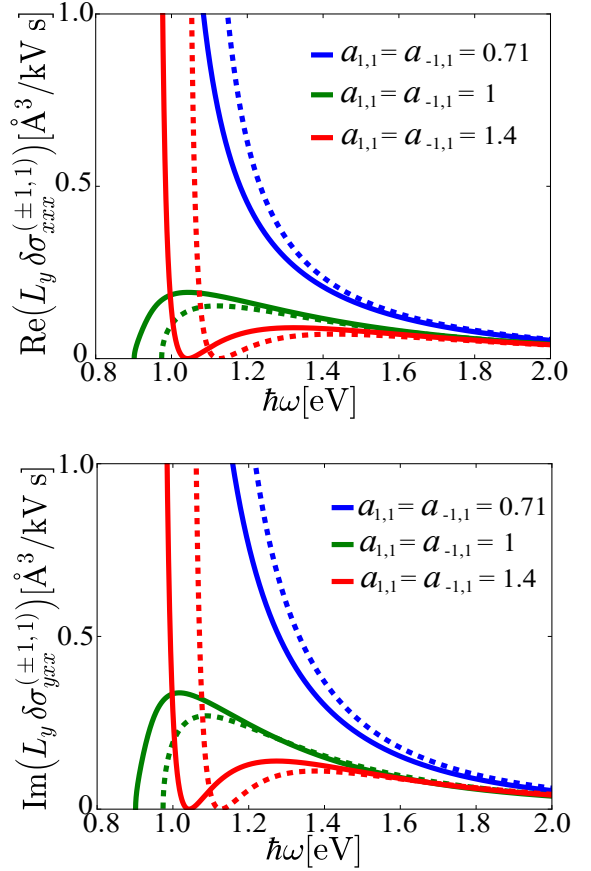}
\caption{\label{Fig:Re_sigma_xxx} Frequency dependence of the resonance terms (\ref{conductivity_resonance_1}), (\ref{conductivity_resonance_2}) in the second order conductivity for different values of the phenomenological parameter in case $a_{1,1}=a_{-1,1}$ in the one valley $K$. Solid (dashed) lines represent contribution of the electrons with the spin \text{$s/2=1/2$} ($s/2=-1/2$). Calculations were carried out at $2\Delta_c=0.148$eV, $2\Delta_v=0.003$eV, $m=0.904$eV, $T=300$K, $\mu=-m+\Delta_v=-0.83$eV, $v=7.7\cdot 10^{5}$m/s.}  
\end{figure}

To conclude, we have shown that the ESs of TMD monolayers possess linear spectra describing by the only phenomenological parameter $a_{s,\tau}$ in absence of valley coupling and spin-orbit interaction at the edge. At \text{$a_{s,1}>0$} $(a_{s,-1}<0)$ the ES spectra intersect the band gap in the $K$($K'$) valley, which leads to resonantly enhanced second harmonic generation in the TMD monolayers. Based on the experimental results on the second harmonic generation from a single layer membranes of MoS$_2$ we have found the value of the phenomenological parameter $a_{1,1}=-a_{-1,-1}=0.71$ for these structures.

This work was partially supported by the Russian Foundation for Basic Research (project 16-32-00655) and the Russian Ministry of Education and Science (''5 top 100'' program).


\appendix*
\begin{widetext}
\section{Derivation of the Eqs.(\ref{conductivity_resonance_1}), (\ref{conductivity_resonance_2})}\label{App:matrix_elements}
Below throughout this section we suppress the indexes $s,\tau$ in $a, \varepsilon_0,\widetilde{m}$ as we are only interested in the matrix elements within the same valley $\tau$ and spin $s$. First, we give an explicit expression for the matrix element $\langle\varepsilon_{1}k_x'|e^{i\bm{qr}}|k_e\rangle$, $\langle k_e|e^{i\bm{qr}}|\varepsilon_{2}k_x\rangle$ in the linear in $\bm{q}$ approximation between the bulk (\ref{bulk_function}) and the edge (\ref{ES_function}) states. They read as follows:
\begin{equation}\label{field_matrix_element_1}
\begin{aligned}
\langle\varepsilon_{1}k_x'|e^{i\bm{qr}}|k_e\rangle \approx \delta_{k_x'k_e}\frac{\sqrt{2}L_xC_eC_{b,1}\left(1-e^{-\kappa L_y}\right)\left(1 - \scalebox{1.2}{$\frac{\tau(\varepsilon_1 - m - s\tau\Delta_c)e^{i\theta}}{\hbar vka}$} \right)}{(\kappa - ik_y)}\times\qquad\qquad\qquad\qquad\qquad\qquad\\
\times\left[iq_x\frac{\tau}{k_y}\left(\scalebox{1.2}{$\frac{\hbar vk_ea-\varepsilon_1 + m + s\tau\Delta_c}{\left(1+a^2\right)\left(\varepsilon_1 +\varepsilon_e - 2\varepsilon_0\right)}$} + \scalebox{1.2}{$\frac{\hbar vk_ea+a^2\left(\varepsilon_1 - m - s\tau\Delta_c\right)}{\left(1+a^2\right)\left(\varepsilon_1 - \varepsilon_e\right)}$} - \scalebox{1.2}{$\frac{\tau\kappa k_e}{\kappa^2 + k_y^2}$}\right) - q_y\scalebox{1.2}{$\frac{\kappa k_y}{\kappa^2 + k_y^2}$}\right]
\end{aligned}
\end{equation}  
\begin{equation}\label{field_matrix_element_2}
\begin{aligned}
\langle k_e|e^{i\bm{qr}}|\varepsilon_{2}k_x\rangle \approx \delta_{k_xk_e}\frac{\sqrt{2}L_xC_eC_{b,2}\left(1-e^{-\kappa L_y}\right)\left(1 - \scalebox{1.2}{$\frac{\tau(\varepsilon_2 - m - s\tau\Delta_c)e^{-i\theta}}{\hbar vka}$} \right)}{(\kappa + ik_y)}\times\qquad\qquad\qquad\qquad\qquad\qquad\\
\times\left[-iq_x\frac{\tau}{k_y}\left(\scalebox{1.2}{$\frac{\hbar vk_ea-\varepsilon_2 + m + s\tau\Delta_c}{\left(1+a^2\right)\left(-\varepsilon_2 + 2\varepsilon_0 -\varepsilon_e \right)}$} + \scalebox{1.2}{$\frac{\hbar vk_ea+a^2\left(\varepsilon_2 - m - s\tau\Delta_c\right)}{\left(1+a^2\right)\left(-\varepsilon_2 + \varepsilon_e\right)}$} + \scalebox{1.2}{$\frac{\tau\kappa k_e}{\kappa^2 + k_y^2}$}\right) + q_y\scalebox{1.2}{$\frac{\kappa k_y}{\kappa^2 + k_y^2}$}\right]
\end{aligned}
\end{equation} 
One can see from Eqs.(\ref{field_matrix_element_1}), (\ref{field_matrix_element_2}) that the terms linear in $q_x$ are proportional to $k_y^{-1}$, but those that linear in $q_y$ are proportional to $k_y$. Therefore, we can infer that the resonance in the second harmonic current will be in terms that are proportional to $E_x^2$. To derive the resonant terms in the non-linear conductivity we also need the matrix elements of the velocity operator between the bulk states (\ref{bulk_function}):
\begin{equation}\label{v_x}
\langle\varepsilon_2k_x|v_x|\varepsilon_1k_x\rangle = \tau v\scalebox{1.2}{$\frac{\hbar v k\left[-\tau\left(1-a^2\right) \hbar vk_x\widetilde{m} + 2\tau a\left(\left(\varepsilon_1 - \varepsilon_{0}\right)^2 - (\hbar v k_x)^2\right)\right]}{\left(\varepsilon_1 - \varepsilon_{0}\right)\left[\left(1-a^2\right)(\hbar v k_x)^2 - 2a\hbar vk_x\widetilde{m} + i2\tau v\hbar k_y\left(\varepsilon_1-\varepsilon_{0}\right)\right]}$}
\end{equation}
\begin{equation}\label{v_y}
\langle\varepsilon_2k_x|v_y|\varepsilon_1k_x\rangle = -iv\scalebox{1.2}{$\frac{\hbar v k\left[\tau\left(1-a^2\right) \hbar vk_x - 2\tau a\widetilde{m}\right]}{\left(1-a^2\right)(\hbar v k_x)^2 - 2a\hbar vk_x\widetilde{m} + i2\tau v\hbar k_y\left(\varepsilon_1 - \varepsilon_{0}\right)}$},
\end{equation}
where we use the identity $\varepsilon_2=-\varepsilon_1 + 2\varepsilon_{0}$ which follows from the equality $k_y(\varepsilon_1,k_e)=k_y(\varepsilon_2,k_e)$. Summation over $\lambda$ in Eq.(\ref{second_density_matrix}),(\ref{current}) implies the following:
\begin{equation}\label{summation}
\sum_{\lambda}\to 
\begin{cases}
\displaystyle
\frac{L_xL_y}{(2\pi )^2}\int dk_xdk_y=\frac{L_xL_y}{(2\pi )^2}\int \frac{\left|\varepsilon_l - \varepsilon_0\right|}{v^2\hbar k_y}d\varepsilon_ldk_x, \mbox{(bulk states)} \\
\\
\displaystyle 
 \frac{L_x}{2\pi}\int dk_x=\frac{L_x(1+a^2)}{4\pi|a|\hbar v}\int d\varepsilon_e, \mbox{(edge states)}
\end{cases}
\end{equation} 
We note that at integration over the bulk states we gain an additional factor $k_y^{-1}$ due to the density of bulk states. Inserting Eqs.(\ref{field_matrix_element_1}-\ref{v_y}) in Eqs.(\ref{current}), (\ref{second_density_matrix}) we obtain that $\delta\sigma_{xyx}^{s\tau} = \delta\sigma_{xxy}^{s\tau} = 0$,
\begin{equation}\label{sigma_xxx}
\begin{aligned}
\delta\sigma_{xxx}^{s\tau} = \tau\frac{e^3v^2\hbar^3}{L_y|a|(2\pi\hbar)^2}\int_{\varepsilon_{e,\text{min}}}^{\varepsilon_{e,\text{max}}} d\varepsilon_{e}\int_{\varepsilon_{1,\text{min}}}^{+\infty} d\varepsilon_{1}\frac{\hbar v\kappa\left[-\tau(1-a^2)v\hbar k_e\widetilde{m}+2\tau a\left((\varepsilon_1-\varepsilon_0)^2-(\hbar vk_e)^2\right)\right]}{(\hbar v k_y)^3\left[(\hbar v\kappa)^2 + (\hbar vk_y)^2\right](\varepsilon_1-\varepsilon_0)\left[2(\varepsilon_1-\varepsilon_0)-2\hbar\omega-i0\right]}\times\qquad\qquad\qquad\qquad\qquad\qquad\qquad\qquad\qquad\qquad\\
\times\left[\scalebox{1.3}{$\frac{\hbar vk_ea-\varepsilon_1 + m + s\tau\Delta_c}{\left(1+a^2\right)\left(\varepsilon_1 +\varepsilon_e - 2\varepsilon_{0}\right)}$} + \scalebox{1.3}{$\frac{\hbar vk_ea+a^2\left(\varepsilon_1 - m - s\tau\Delta_c\right)}{\left(1+a^2\right)\left(\varepsilon_1 - \varepsilon_e\right)}$} - \scalebox{1.3}{$\frac{\tau\kappa k_e}{\kappa^2 + k_y^2}$}\right]^2\left[\frac{f(-\varepsilon_1+2\varepsilon_0) - f(\varepsilon_e)}{-\varepsilon_1+2\varepsilon_0 - \varepsilon_e + \hbar\omega + i0} - \frac{f(\varepsilon_e) - f(\varepsilon_1)}{\varepsilon_e - \varepsilon_1 + \hbar\omega + i0}\right],\qquad\qquad\qquad\qquad\qquad\qquad\qquad\qquad
\end{aligned}
\end{equation}
\begin{equation}\label{sigma_yxx}
\begin{aligned}
\delta\sigma_{yxx}^{s\tau} = \frac{-ie^3v^2\hbar^3}{L_y|a|(2\pi\hbar)^2}\int_{\varepsilon_{e,\text{min}}}^{\varepsilon_{e,\text{max}}} d\varepsilon_{e}\int_{\varepsilon_{1,\text{min}}}^{+\infty}d\varepsilon_{1}\frac{\hbar v\kappa\left[\tau(1-a^2)v\hbar k_e-2\tau a\widetilde{m}\right]}{(\hbar vk_y)^3\left[(\hbar v\kappa)^2 + (\hbar vk_y)^2\right]\left[2(\varepsilon_1-\varepsilon_0)-2\hbar\omega-i0\right]}\times\qquad\qquad\qquad\qquad\qquad\qquad\qquad\qquad\qquad\qquad\\
\times\left[\scalebox{1.3}{$\frac{\hbar vk_ea-\varepsilon_1 + m + s\tau\Delta_c}{\left(1+a^2\right)\left(\varepsilon_1 +\varepsilon_e - 2\varepsilon_{0}\right)}$} + \scalebox{1.3}{$\frac{\hbar vk_ea+a^2\left(\varepsilon_1 - m - s\tau\Delta_c\right)}{\left(1+a^2\right)\left(\varepsilon_1 - \varepsilon_e\right)}$} - \scalebox{1.3}{$\frac{\tau\kappa k_e}{\kappa^2 + k_y^2}$}\right]^2\left[\frac{f(-\varepsilon_1+2\varepsilon_0) - f(\varepsilon_e)}{-\varepsilon_1+2\varepsilon_0 - \varepsilon_e + \hbar\omega + i0} - \frac{f(\varepsilon_e) - f(\varepsilon_1)}{\varepsilon_e - \varepsilon_1 + \hbar\omega + i0}\right],\qquad\qquad\qquad\qquad\qquad\qquad
\end{aligned}
\end{equation}
\begin{equation}\label{sigma_yyy}
\begin{aligned}
\delta\sigma_{yyy}^{s\tau} = \frac{ie^3v^2\hbar^3}{L_y|a|(2\pi\hbar)^2}\int_{\varepsilon_{e,\text{min}}}^{\varepsilon_{e,\text{max}}} d\varepsilon_{e}\int_{\varepsilon_{1,\text{min}}}^{+\infty}d\varepsilon_{1}\frac{(\hbar v\kappa)(\hbar vk_y)\left[\tau(1-a^2)v\hbar k_e-2\tau a\widetilde{m}\right]}{\left[(\hbar v\kappa)^2 + (\hbar vk_y)^2\right]^3\left[2(\varepsilon_1-\varepsilon_0)-2\hbar\omega-i0\right]}\times\\
\times\left[\frac{f(-\varepsilon_1+2\varepsilon_0) - f(\varepsilon_e)}{-\varepsilon_1+2\varepsilon_0 - \varepsilon_e + \hbar\omega + i0} - \frac{f(\varepsilon_e) - f(\varepsilon_1)}{\varepsilon_e - \varepsilon_1 + \hbar\omega + i0}\right],
\end{aligned}
\end{equation}
\begin{equation}\label{sigma_xyy}
\begin{aligned}
\delta\sigma_{xyy}^{s\tau} = -\tau\frac{e^3v^2\hbar^3}{L_y|a|(2\pi\hbar)^2}\int_{\varepsilon_{e,\text{min}}}^{\varepsilon_{e,\text{max}}} d\varepsilon_{e}\int_{\varepsilon_{1,\text{min}}}^{+\infty}d\varepsilon_{1}\frac{(\hbar v\kappa)(\hbar v k_y)\left[-\tau(1-a^2)v\hbar k_e\widetilde{m}+2\tau a\left((\varepsilon_1-\varepsilon_0)^2-(\hbar vk_e)^2\right)\right]}{\left[(\hbar v\kappa)^2 + (\hbar vk_y)^2\right]^3|\varepsilon_1-\varepsilon_0|\left[2(\varepsilon_1-\varepsilon_0)-2\hbar\omega-i0\right]}\times\qquad\qquad\qquad\qquad\qquad\qquad\qquad\qquad\qquad\\
\times\left[\frac{f(-\varepsilon_1+2\varepsilon_0) - f(\varepsilon_e)}{-\varepsilon_1+2\varepsilon_0 - \varepsilon_e + \hbar\omega + i0} - \frac{f(\varepsilon_e) - f(\varepsilon_1)}{\varepsilon_e - \varepsilon_1 + \hbar\omega + i0}\right],\qquad\qquad\qquad\qquad\qquad\qquad\qquad\qquad\qquad\qquad\qquad\qquad\qquad
\end{aligned}
\end{equation}
where the integration limits are the following \text{$\varepsilon_{1,\text{min}}=\varepsilon_0+\left\{\widetilde{m}^2 + (\hbar vk_e)^2\right\}^{1/2}$}, and 
$
\begin{cases}
\displaystyle
\varepsilon_{e,\text{min}}=-\infty$, $\varepsilon_{e,\text{max}}=-m\frac{1+a^2}{1-a^2}+\frac{s\tau\left(\Delta_v-a^2\Delta_c\right)}{1-a^2},\quad \text{at} \quad 0<a<1 \\ 
\varepsilon_{e,\text{min}}=-m\frac{1+a^2}{1-a^2}+\frac{s\tau\left(\Delta_v-a^2\Delta_c\right)}{1-a^2}$, $\varepsilon_{e,\text{max}}=+\infty,\quad \text{at} \quad a>1
\end{cases}
$

On the final step using the Sokhotski–-Plemelj formula in Eqs.(\ref{sigma_xxx}--\ref{sigma_xyy})
\begin{equation}
\frac{1}{x\pm i0} = \mp i\pi\delta(x) + \text{P}\frac{1}{x}
\end{equation} 
we arrive to the following formulae for contributions which contain the product of the two Dirac delta-functions in the integrand functions:
\begin{equation}\label{Re_sigma_xxx}
\begin{aligned}
\displaystyle
{\rm Re}\left(\delta\sigma^{(s,\tau)}_{xxx}\right) = \frac{\tau e^3v^2}{4L_y(\hbar\omega)^2}\frac{\omega_{0,s\tau}\left[\omega - \scalebox{1.2}{$\frac{2|a_{s\tau}|}{1+a_{s\tau}^2}$}\omega_{0,s\tau}+\scalebox{1.2}{$\frac{\left(1-a_{s\tau}^2\right)\omega_{0,s\tau}^2}{2\left(1+a_{s\tau}^2\right)\omega}$}\right]^2}{\omega\left[\omega^2 - \omega_{0,s\tau}^2\right]^{3/2}}\Theta(\omega-\omega_{0,s\tau})\times\\
\times\left[f\left(\varepsilon_{0,s,\tau}-\hbar\omega\right)-2f\left(\varepsilon_{0,s,\tau}\right)+f\left(\varepsilon_{0,s,\tau}+\hbar\omega\right)\right]\qquad
\end{aligned}
\end{equation}
\begin{equation}\label{Im_sigma_yxx}
\begin{aligned}
{\rm Im}\left(\delta\sigma^{(s,\tau)}_{yxx}\right) = \frac{e^3v^2\left(1+a_{s\tau}^2\right)}{4L_y\left|a_{s\tau}\right|(\hbar\omega)^2}\frac{\omega_{0,s\tau}^2\left[\omega - \scalebox{1.2}{$\frac{2|a_{s\tau}|}{1+a_{s\tau}^2}$}\omega_{0,s\tau}+\scalebox{1.2}{$\frac{\left(1-a_{s\tau}^2\right)\omega_{0,s\tau}^2}{2\left(1+a_{s\tau}^2\right)\omega}$}\right]^2}{\omega^2\left[\omega^2 - \omega_{0,s\tau}^2\right]^{3/2}}\Theta(\omega-\omega_{0,s\tau})\times\qquad\\
\times\left[f\left(\varepsilon_{0,s,\tau}-\hbar\omega\right)-2f\left(\varepsilon_{0,s,\tau}\right)+f\left(\varepsilon_{0,s,\tau}+\hbar\omega\right)\right]
\end{aligned}
\end{equation}
\begin{equation}\label{Im_sigma_yyy}
{\rm Im}\left(\delta\sigma_{yyy}^{(s,\tau)}\right) =-\frac{ie^3v^2(1+a^2)\omega_0^2\sqrt{\omega^2-\omega_0^2}}{8L_y|a|(\hbar\omega)^2\omega^4}\Theta(\omega-\omega_{0,s\tau})\left[f\left(\varepsilon_{0,s,\tau}-\hbar\omega\right)-2f\left(\varepsilon_{0,s,\tau}\right)+f\left(\varepsilon_{0,s,\tau}+\hbar\omega\right)\right]
\end{equation}
\begin{equation}\label{Re_sigma_xyy}
{\rm Re}\left(\delta\sigma_{xyy}^{(s,\tau)}\right) =-\frac{\tau e^3v^2\omega_0\sqrt{\omega^2-\omega_0^2}}{4L_y|a|(\hbar\omega)^2\omega^3}\Theta(\omega-\omega_{0,s\tau})\left[f\left(\varepsilon_{0,s,\tau}-\hbar\omega\right)-2f\left(\varepsilon_{0,s,\tau}\right)+f\left(\varepsilon_{0,s,\tau}+\hbar\omega\right)\right]
\end{equation}
The power-law singularity $\left(\omega - \omega_0\right)^{-3/2}$ emerges only in those components of the non-linear conductivity where the integrand function in Eqs.(\ref{sigma_xxx}--\ref{sigma_xyy}) is proportional to $k_y^{-3}$. 
\end{widetext}

\end{document}